\documentclass[conference,letterpaper]{IEEEtran}
\ifCLASSINFOpdf
\else
\fi
\usepackage{xcolor}
\usepackage{graphicx}
\usepackage{listings}
\usepackage[most]{tcolorbox}
\usepackage{longtable} 
\usepackage{array} 
\usepackage{tabularx}
\usepackage{booktabs}
\usepackage{caption}
\usepackage{subcaption}
\usepackage{algorithm}
\usepackage{algorithmic}
\usepackage{amsmath}
\usepackage{amssymb}
\begin{document}
\pagestyle{empty}
\title{LLM-Based Threat Detection and Prevention Framework for IoT Ecosystems}
%
%
%

\author{
\IEEEauthorblockN{
Yazan Otoum\IEEEauthorrefmark{1},
Arghavan Asad\IEEEauthorrefmark{1},
Ishtiaq Ahmad\IEEEauthorrefmark{3},
Amiya Nayak\IEEEauthorrefmark{2}}
\vspace{-0.1em}

\IEEEauthorblockA{\small \IEEEauthorrefmark{1}School of Computer Science and Technology, Algoma University, Canada}
\vspace{-0.2em}

\IEEEauthorblockA{\small \IEEEauthorrefmark{2}School of Electrical Engineering and Computer Science, University of Ottawa, Canada}
\vspace{-0.2em}

\IEEEauthorblockA{\small \IEEEauthorrefmark{3}Faculty of Electrical Engineering, Czech Technical University in Prague, Prague, Czech Republic and \\ School of Electrical Engineering and Computer Science (SEECS), National University of Sciences and Technology (NUST), Islamabad}
}


\maketitle

\pagenumbering{gobble}

\begin{abstract}
The rapid expansion of the Internet of Things (IoT) has introduced complex and evolving cybersecurity challenges. This paper presents a Large Language Model (LLM)-based framework for intelligent threat detection and prevention in IoT networks. The proposed architecture integrates two core components: a Detection Component that leverages fine-tuned models trained on IoT-specific datasets (IoT-23 and TON\_IoT) to identify anomalous network behaviours, and a Prevention Component that generates adaptive response strategies to mitigate detected threats. The framework incorporates prompt management, local adaptation, and global refinement modules to support context-aware decision-making and continuous model-guided adaptation. Experimental validation in a Docker-based edge--cloud environment shows that BERT Small achieves 99.75\% accuracy, F1-score, precision, and recall for threat detection, while the Prevention LLM achieves 98.3\% accuracy in selecting suitable mitigation actions. These results highlight the potential of LLM-driven architectures to provide explainable and adaptive security solutions for next-generation IoT ecosystems.
\end{abstract}


%
\IEEEpeerreviewmaketitle

\section{Introduction}

\IEEEPARstart{T}{he} rapid growth of Internet of Things (IoT) ecosystems has introduced significant challenges in managing, optimizing, and securing system performance. One of the primary drivers of these challenges is the massive volume of data generated by IoT environments, which requires efficient, real-time processing and advanced operations such as resource allocation, predictive maintenance, and energy management. These challenges are further intensified by the diverse and dynamic nature of IoT devices, which often operate under heterogeneous conditions with varying computational and network capabilities \cite{zafar2023advanced}. In addition, the large-scale expansion of IoT networks, now comprising billions of connected devices, has increased the complexity of maintaining secure and reliable distributed systems. Traditional IoT security techniques often struggle to detect evolving and sophisticated threats in such dynamic environments \cite{dritsas2026deployment}. 

Several studies have demonstrated the potential of Machine Learning (ML) algorithms to identify unusual patterns in IoT network traffic and to provide appropriate responses to emerging threats. For instance, the work~\cite{krzyszton2024comparative} presented a comprehensive study on the use of ML for anomaly detection in IoT systems, emphasizing its ability to accurately detect unauthorized access and other security breaches. The authors of \cite{chellammal2022fast} expanded on this by proposing a multi-layered machine learning approach for intrusion detection that significantly outperformed traditional rule-based systems, achieving better identification of complex attack patterns and reducing false positives. Beyond conventional ML techniques, deep learning models such as Convolutional Neural Networks (CNNs) and Recurrent Neural Networks (RNNs) have been extensively utilized for intrusion detection in IoT networks. For example, the authors of~\cite{hindy2020utilising} proposed an autoencoder-based deep learning model that effectively identifies zero-day attacks while maintaining a low false-positive rate. Another significant work \cite{jamshidi2025application} integrated reinforcement learning with anomaly detection, allowing IoT security systems to adapt dynamically to new attack patterns without extensive retraining. 

Recent advances in Large Language Models (LLMs) have shown strong potential to enhance IoT security systems by providing adaptability, scalability, and effectiveness in handling complex datasets. The authors of \cite{worae2024unified} introduced a LLM-based approach to network traffic analysis, using prompt engineering to enhance the model's ability to detect malicious activity in IoT networks. Their findings suggested that LLMs could achieve results comparable to deep learning models, but with reduced computational overhead and improved resource efficiency. Another critical advancement in IoT security is the adoption of Federated Learning (FL) for distributed threat detection. Since IoT devices generate vast amounts of decentralized data, transmitting all traffic logs to a centralized server introduces privacy risks and scalability challenges. FL addresses this by enabling devices to collaboratively train models without sharing raw data \cite{otoum2025llms}. \cite{divakaran2024llms} investigated domain-specific fine-tuning of LLMs to identify zero-day vulnerabilities. Similarly, \cite{chen2021learning} proposed a hybrid framework that combines LLMs with graph-based anomaly detection to secure IoT networks against emerging threats. The integration of LLMs with FL frameworks has also been shown to enhance data privacy while maintaining robust security in distributed IoT environments \cite{adjewa2024efficient}. Moreover, \cite{koli2025ai} highlighted the effectiveness of context-aware LLMs in identifying complex attack patterns specific to IoT ecosystems, while \cite{hasan2024distributed} emphasized the importance of adaptive learning techniques for continuous improvement in threat detection and prevention. Motivated by these advances, our work presents an LLM-based security system tailored for IoT environments using a two-component framework for threat detection and prevention. The proposed system leverages prompt engineering, fine-tuning, and domain-specific training to identify and respond to security breaches in real time. Experimental evaluation compares lightweight Bidirectional Encoder Representations from Transformers (BERT)-based variants within the proposed framework to determine the most effective model for the detection component. The main contributions of this work are:
\begin{itemize}
    \item We propose an edge--cloud LLM-based framework for IoT threat detection and prevention, integrating BERT-based detection, LLM-driven prevention and cloud-assisted refinement.

    \item We design a context-aware prevention mechanism that uses Mistral-7B-v0.1 to generate structured mitigation actions, which are validated by the Local Adaptation Engine before execution to ensure policy compliance and operational suitability.

    \item We implement a reproducible Docker-based edge--cloud testbed that emulates IoT devices, edge nodes, and cloud services for evaluating detection and prevention.

    \item We validate the framework using IoT-23 and TON\_IoT, where BERT Small achieves 99.75\% accuracy, F1-score, precision, and recall, and the Prevention LLM achieves 98.3\% accuracy in selecting suitable mitigation actions.
\end{itemize}

\section{System Model and Problem Formulation}
\label{sec:SystemModel}
This section defines the system environment, data flow, and security objectives, outlining the roles of the IoT, edge, and cloud layers. It also formulates the core problem: classifying IoT traffic and selecting appropriate mitigation actions in an edge--cloud setting using observed features and historical feedback.

\subsection{System Model}

We consider an edge--cloud IoT security architecture consisting of heterogeneous IoT devices, an edge gateway, and cloud services. The IoT environment includes devices such as smart cameras, industrial sensors, smart meters, wearable health sensors, and smart home devices. The edge gateway performs data collection, preprocessing, and local security coordination. IoT devices generate traffic logs, flow-level features, system events, and device-context information. Let $\mathcal{D} = \{d_1, d_2, \ldots, d_N\}$ denote the set of IoT devices connected to the edge gateway, where $N$ is the number of devices. Let $M$ denote the total number of observations and $p$ denote the number of features per observation. The observation set is defined as $\mathcal{X} = \{x_1, x_2, \ldots, x_M\}$, where each $x_j \in \mathbb{R}^p$ represents traffic-related and context-related information collected from the IoT environment. Thus, the full dataset is represented as $X \in \mathbb{R}^{M \times p}$, where each row is an observation and each column is a feature.

\subsection{Problem Formulation}
The objective is to design an intelligent IoT security framework that jointly optimizes threat detection and adaptive mitigation to maximize detection accuracy and prevention effectiveness in dynamic IoT environments. Given the observation set $\mathcal{X}$, where each $x_j \in \mathbb{R}^p$ is defined in the system model and generated by IoT devices in $\mathcal{D}$, the detection component predicts the traffic class and confidence score as $f_{\theta}(x_j) = (y_j, c_j)$, where $y_j \in \mathcal{Y}$ and $c_j \in [0,1]$, and $\mathcal{Y}$ denotes the set of benign and attack classes. For observations classified as malicious, the prevention component generates a mitigation action given by $a_j = g_{\phi}(x_j, y_j, c_j, m_j, h_j)$, where $a_j \in \mathcal{A}$, $m_j$ represents contextual metadata, and $h_j$ denotes historical feedback. The optimization problem is defined as
\begin{align}
    f_{\theta}^{*}, g_{\phi}^{*} =
    \arg\max_{f_{\theta},\,g_{\phi}}
    & \;\; \alpha F1(f_{\theta}) + \beta Recall(f_{\theta}) + \gamma Acc(g_{\phi}) \nonumber \\
    \text{s.t.} \quad
    & \text{(a)}\;\; a_j \in \mathcal{A}, \quad \forall j.
\end{align}
where $F1(f_{\theta})$ and $Recall(f_{\theta})$ evaluate detection performance under imbalanced traffic classes, $Acc(g_{\phi})$ denotes the accuracy of the Prevention Component in selecting suitable mitigation actions, and $\alpha$, $\beta$, and $\gamma$ are weighting factors. Constraint (a) ensures that each selected mitigation action lies within the feasible action space $\mathcal{A}$, which comprises valid mitigation responses in IoT security environments.

\section{Proposed Framework}
\label{sec:ProposedFramework}

\subsection{Architecture Overview}

\begin{figure*}[t]
    \centering
\includegraphics[width=0.9\textwidth]{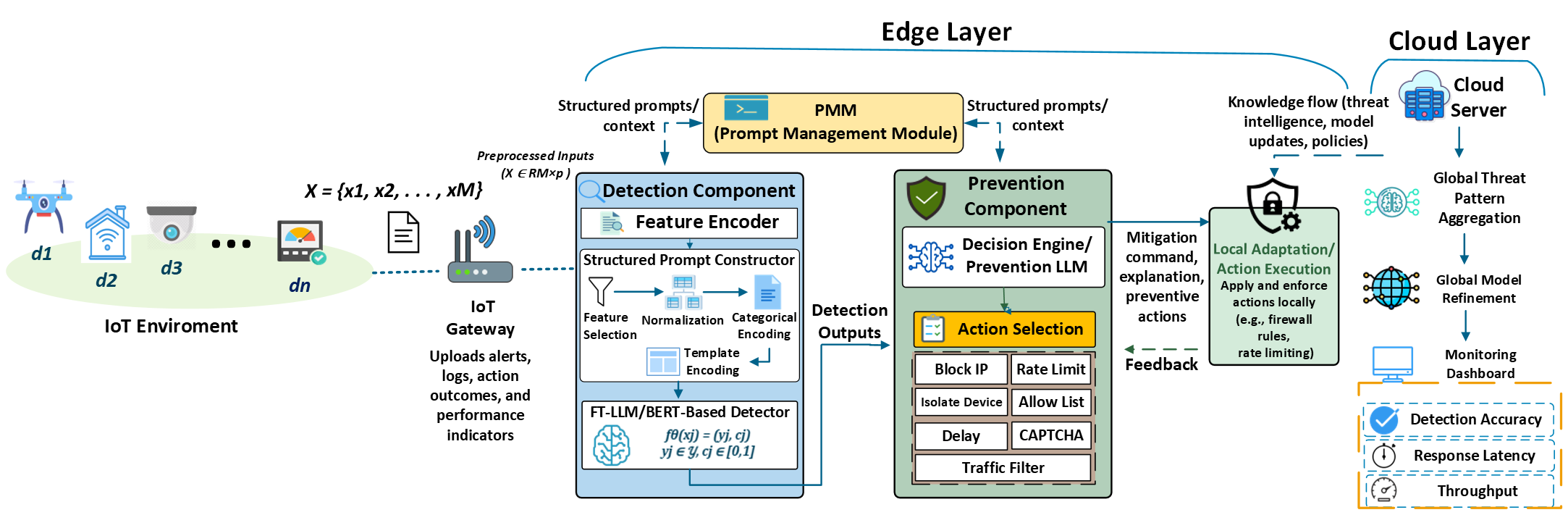}
\caption{Overview of the proposed framework.}
\label{fig:Architecture}
\end{figure*}

The proposed framework integrates threat detection and prevention within an edge--cloud IoT architecture, as shown in Fig.~\ref{fig:Architecture}. Given the observation set $\mathcal{X}$, where each $x_j \in \mathbb{R}^p$ is a traffic feature vector, the edge layer performs preprocessing, detection, prevention decision-making, and action execution, while the cloud layer supports monitoring, threat aggregation, and prompt-based refinement. The Detection Component implements $f_{\theta}(x_j)=(y_j,c_j)$ using BERT Small, selected for its balanced accuracy and inference performance on IoT-23 and TON\_IoT. The PMM serves as a shared module that prepares structured prompts, contextual fields, and few-shot exemplars for both Detection and Prevention Components. The Prevention Component implements $a_j=g_{\phi}(x_j,y_j,c_j,m_j,h_j)$ using Mistral-7B-v0.1, where $a_j \in \mathcal{A}$ is the selected mitigation action. Before execution, the Local Adaptation Engine validates each action for feasibility, policy compliance, and suitability for the device or traffic. At the cloud layer, prompts, exemplars, policy rules, and threat-intelligence records are updated based on aggregated alerts, logs, outcomes, and performance indicators without retraining the model weights.

\subsection{Detection Component} 
The Detection Component identifies anomalous behaviours and potential security breaches within IoT networks by implementing the detection function $f_{\theta}(x_j) = (y_j, c_j)$, where each input observation $x_j \in \mathbb{R}^p$ represents a traffic feature vector and the output consists of a predicted traffic class $y_j \in \mathcal{Y}$ and a confidence score $c_j \in [0,1]$. The model is based on a fine-tuned BERT architecture trained on IoT-specific datasets, including IoT-23 and TON\_IoT. These datasets are selected because they contain realistic IoT network traffic, multiple attack categories, and flow-level records suitable for intrusion detection. Before training, the datasets are merged into a unified corpus, their labels are harmonized into a common multi-class classification space $\mathcal{Y}$, and the traffic observations are preprocessed into normalized feature representations. The detector performs multi-class classification rather than binary benign/malicious detection. Each observation $x_j$ is constructed from traffic-related and context-related features, including protocol information, service type, connection duration, packet and byte statistics, connection state, source/destination ports, and device-level metadata when available. This representation enables the model to capture both traffic patterns and operational context. The prompt-construction process follows three steps. First, relevant features are selected from each observation $x_j$. Second, numerical fields are normalized and categorical fields (e.g., protocol, service, connection state) are encoded into structured textual representations. Third, the selected features are inserted into a fixed prompt template to produce a compact and structured input for inference. The output of the Detection Component includes the predicted traffic class $y_j$, confidence score $c_j$, severity level, and, when applicable, attacker-related metadata. This output is forwarded to the Prevention Component, which uses $(x_j, y_j, c_j)$ and contextual metadata to generate a mitigation action, and to the Local Adaptation Engine for validation and enforcement. The Detection Component is designed for deployment on computationally capable edge infrastructure, such as IoT gateways, edge servers, or local security orchestration nodes. In this work, BERT Small is selected as the primary detection model due to its favourable trade-off between detection accuracy and inference efficiency among the evaluated lightweight BERT-based models. Cloud-assisted refinement supports the Detection Component through periodic updates to prompt templates, few-shot exemplars, policy rules, and threat-intelligence records. These updates are derived from aggregated alerts, logs, action outcomes, and monitoring feedback collected from the edge layer, enabling continuous improvement of the detection process without modifying model weights during operation. The selected features are commonly used in network intrusion detection because they capture connection behaviour, communication volume, protocol activity, and session state. For example, the connection state ``SF'' indicates a connection that was successfully established and normally terminated. The following example shows a structured input representation of an observation $x_j$ and the corresponding prediction $(y_j, c_j)$ generated by the detection component:

\begin{tcolorbox}[
colback=gray!10,
colframe=gray!50,
sharp corners,
boxrule=0.3pt,
enhanced,
boxsep=1pt,
left=2pt,
right=2pt,
top=2pt,
bottom=2pt,
before skip=2pt,
after skip=2pt
]
\scriptsize
\textbf{Input:}

Traffic from Port 49864 to Port 80 over TCP Protocol. Duration: 0.049751s, Service: http, Bytes Sent: 243, Bytes Received: 3440, Missed Bytes: 0, Total IP Bytes Sent: 511, Total IP Bytes Received: 3760, Packets Sent: 5, Packets Received: 6, Connection State: SF.

\textbf{Prediction:}

Traffic Class: DDoS. Confidence: High. Severity: High.
\end{tcolorbox}
\subsection{Prevention Component} 
The Prevention Component implements the decision function $a_j = g_{\phi}(x_j, y_j, c_j, m_j, h_j)$ to generate adaptive mitigation actions based on the Detection Component's outputs. The input includes the predicted traffic class $y_j$, confidence score $c_j$, and contextual metadata $m_j$, while $h_j$ represents historical mitigation actions and their outcomes. The contextual metadata is derived from monitored traffic and system state, including attacker identifiers (e.g., IP address or device ID), system load, and traffic characteristics. Attack intensity is estimated from indicators such as packet rate, failed login attempts, traffic volume, and repeated alerts within a time window. Source distribution is inferred from the number and frequency of suspicious sources, while system load reflects the current resource utilization of the affected edge node. These contextual features are structured and inserted into predefined prompt templates by the PMM. The Prevention Component selects a mitigation action $a_j \in \mathcal{A}$, where $\mathcal{A}$ denotes the set of feasible response strategies for IoT security environments. Since not all actions are valid for every device or attack type, the selected action is validated by the Local Adaptation Engine before execution. The Local Adaptation Engine performs validation through syntax checking, policy compliance, whitelist rules, device criticality, and resource availability checks. Actions that violate local policies or may disrupt legitimate services are rejected or replaced with safer alternatives. This design enables adaptive response generation, as the selected action $a_j$ depends on the current observation $x_j$, the detection output $(y_j, c_j)$, the contextual metadata $m_j$, and the historical feedback $h_j$. It also improves interpretability, as each generated action is accompanied by a short rationale that is logged for monitoring and further analysis. The following examples illustrate the structured prompt format corresponding to the input $(x_j, y_j, m_j, h_j)$ and the resulting mitigation action $a_j$ generated by the Prevention Component:

\begin{tcolorbox}[
colback=gray!10,
colframe=gray!50,
sharp corners,
boxrule=0.3pt,
enhanced,
boxsep=1pt,
left=2pt,
right=2pt,
top=2pt,
bottom=2pt,
before skip=2pt,
after skip=2pt
]
\scriptsize
\textbf{Prompt:}

\texttt{Attack Type:} DDoS\\
\texttt{Attacker IP:} 110.143.175.79\\
\texttt{Intensity:} High\\
\texttt{Duration:} Short\\
\texttt{System Load:} Low\\
\texttt{Previous Status for this IP:}\\
\hspace*{1em}Intensity: Moderate\\
\hspace*{1em}Duration: Short\\
\hspace*{1em}System Load: Low\\
\texttt{Previous Action:} Rate limit IP\\
\texttt{Effectiveness:} Ineffective\\
\texttt{Suggest new prevention action.}

\textbf{Answer:}

Block IP. Explanation: The attack has escalated to high intensity, and rate limiting was ineffective. Blocking the IP is the necessary next step.
\end{tcolorbox}

\subsection{Prompt Management Module}
The PMM is a shared support module for the Detection and Prevention Components. At the edge layer, it prepares structured prompts using task-specific templates, contextual fields, and few-shot exemplars. For each detection, prevention, prompt-update, or monitoring request, the PMM identifies the task type from metadata, such as the source module, alert type, detected attack class, and requested operation. It then retrieves the appropriate template and exemplars from a local repository using rule-based matching based on threat category, device type, and the latest validated prompt version. For detection, the PMM converts normalized traffic features into structured inputs for the BERT-based detector. For prevention, it combines detection outputs, contextual metadata, action history, and policy constraints into prompts for the Prevention LLM. At the cloud layer, the PMM maintains and synchronizes versioned prompt templates, few-shot exemplars, and policy rules derived from alerts, logs, action outcomes, and monitoring feedback. If an edge node is offline, it uses the latest cached prompt version and receives updates once connectivity is restored.
\subsection{Local Adaptation Engine}
The Local Adaptation Engine is an edge-level module that validates and enforces mitigation actions generated by the Prevention Component. It receives the detection output and candidate action $a_j$, then checks whether the action is feasible, policy-compliant, and suitable for the affected device or traffic condition. The engine selects the most suitable mitigation strategy based on expected effectiveness, device criticality, system load, and service-disruption risk. Device criticality is defined by the device's operational role, such as safety-sensitive, health-related, industrial, or routine monitoring. Before execution, the engine applies post-validation checks, including confidence-threshold verification, whitelist matching, resource availability, and action compatibility. Actions that violate local policies or may disrupt legitimate services are rejected or replaced with safer alternatives. After execution, action outcomes and performance indicators are reported to the Monitoring module and cloud layer for logging and future refinement.
\subsection{Monitoring}
The Monitoring module is a cloud-supported component that tracks edge-layer evaluation and operational metrics. During offline evaluation, it computes accuracy, precision, recall, and F1-score using labelled data. During online operation, where ground truth may be unavailable, it monitors alert frequency, confidence scores, selected actions, outcomes, throughput, energy use, resource utilization, and system availability. These metrics are visualized through a cloud dashboard and can trigger operator review, alert escalation, prompt updates, threshold tuning, or policy refinement. The module also supports the edge--cloud feedback loop by forwarding logs, alerts, and performance data to the Global Threat Pattern Aggregation and Global Model Refinement modules. Communication uses authenticated and encrypted channels, and edge nodes continue operating with cached configurations during temporary connectivity loss.
\subsection{Global Threat Pattern Aggregation}
The Global Threat Pattern Aggregation module is a cloud-level service that analyzes alerts, logs, and action outcomes from multiple edge nodes while local detection and response continue at the edge. It correlates structured metadata, including anonymized source identifiers, predicted attack classes, device types, confidence scores, outcomes, and timestamps, using fixed or sliding time windows. Sensitive identifiers, such as IP addresses or device IDs, can be anonymized or hashed before cloud-side analysis. Rule-based correlation identifies recurring-source patterns, such as repeated high-confidence alerts from the same source, and coordinated-threat patterns, such as similar attacks reported across multiple nodes. The resulting insights are forwarded to the Global Model Refinement module to update prompts, exemplars, policies, and threat-intelligence records.
\subsection{Global Model Refinement}
The Global Model Refinement module is a cloud-based engine that updates the prompt and policy resources used by edge PMMs without retraining model weights. It supports both Detection and Prevention components by refining prompt templates, few-shot exemplars, policy rules, calibration guidance, and threat intelligence. When recurring threats are identified, representative cases are selected based on confidence scores, mitigation outcomes, deduplication, and policy checks, then converted into version-controlled prompt packs. Each pack is validated for format, policy compliance, action correctness, and consistency before being securely distributed to edge PMMs via authenticated and encrypted channels. This process reduces overhead while enabling safe and adaptive refinement across edge nodes.

\begin{table*}[t]
    \centering
    \vspace*{0.07in}
    \caption{Performance Metrics and Resource Usage for the LLMs}
    \label{tab:results}
    \renewcommand{\arraystretch}{1.2}
    \begin{tabular}{|>{\centering\arraybackslash}p{2cm}|
                        >{\centering\arraybackslash}p{1cm}|
                        >{\centering\arraybackslash}p{1cm}|
                        >{\centering\arraybackslash}p{1.3cm}|
                        >{\centering\arraybackslash}p{1.3cm}|
                        >{\centering\arraybackslash}p{1.3cm}|
                        >{\centering\arraybackslash}p{1.2cm}|
                        >{\centering\arraybackslash}p{1.3cm}|
                        >{\centering\arraybackslash}p{1.5cm}|
                        >{\centering\arraybackslash}p{1.5cm}|}
        \hline
        \textbf{Model} & 
        \textbf{Param.} & 
        \textbf{Train Loss} & \textbf{Validation Loss} & \textbf{Test Accuracy} & \textbf{Test F1-Score} & \textbf{Test Precision} & \textbf{Test Recall} & \textbf{Energy Consumption (J/Req)} & \textbf{Inference (Req/Sec)} \\ 
        \hline
        BERT Small \cite{bertsmall1, bertsmall2} & 29.1M & 0.0089 & 0.0086 & 99.75\% & 99.75\% & 99.75\% & 99.75\% & 0.1434 & 287.82 \\ 
        \hline
        BERT Mini \cite{bertsmall1, bertsmall2}  & 11.3M & 0.0104 & 0.0097 & 99.72\% & 99.72\% & 99.72\% & 99.72\% & 0.1175 & 293.56 \\ 
        \hline
        TinyBERT \cite{tinybert} & 14.5M & 0.0098  & 0.0096 & 99.73\% & 99.73\% & 99.73\% & 99.73\% & 0.1522 & 232.35 \\ 
        \hline
    \end{tabular}
\end{table*}

\begin{table*}[h]
    \centering
    \caption{Performance Metrics and Resource Usage for the Prevention LLM}
    \label{tab:results_prevention}
    \renewcommand{\arraystretch}{1.2}
    \begin{tabular}{|>{\centering\arraybackslash}p{2.5cm}|
                        >{\centering\arraybackslash}p{2cm}|
                        >{\centering\arraybackslash}p{1.5cm}|
                        >{\centering\arraybackslash}p{1.3cm}|
                        >{\centering\arraybackslash}p{1.3cm}|
                        >{\centering\arraybackslash}p{2.5cm}|
                        >{\centering\arraybackslash}p{1.5cm}|}
        \hline
        \textbf{Model} & 
        \textbf{Parameters} & 
        \textbf{Train Loss} & 
        \textbf{Validation Loss} & 
        \textbf{Test Accuracy} & 
        \textbf{Energy Consumption (J/Req)} & 
        \textbf{Inference (Req/Sec)} \\ 
        \hline
        Mistral-7B-v0.1 \cite{mistral7b} & 7B & 0.0037 & 0.0066 & 98.3\% & 126.46 & 0.46 \\ 
        \hline
    \end{tabular}
\end{table*}

\begin{figure}[t]
    \centering
    \includegraphics[width=0.4\textwidth]{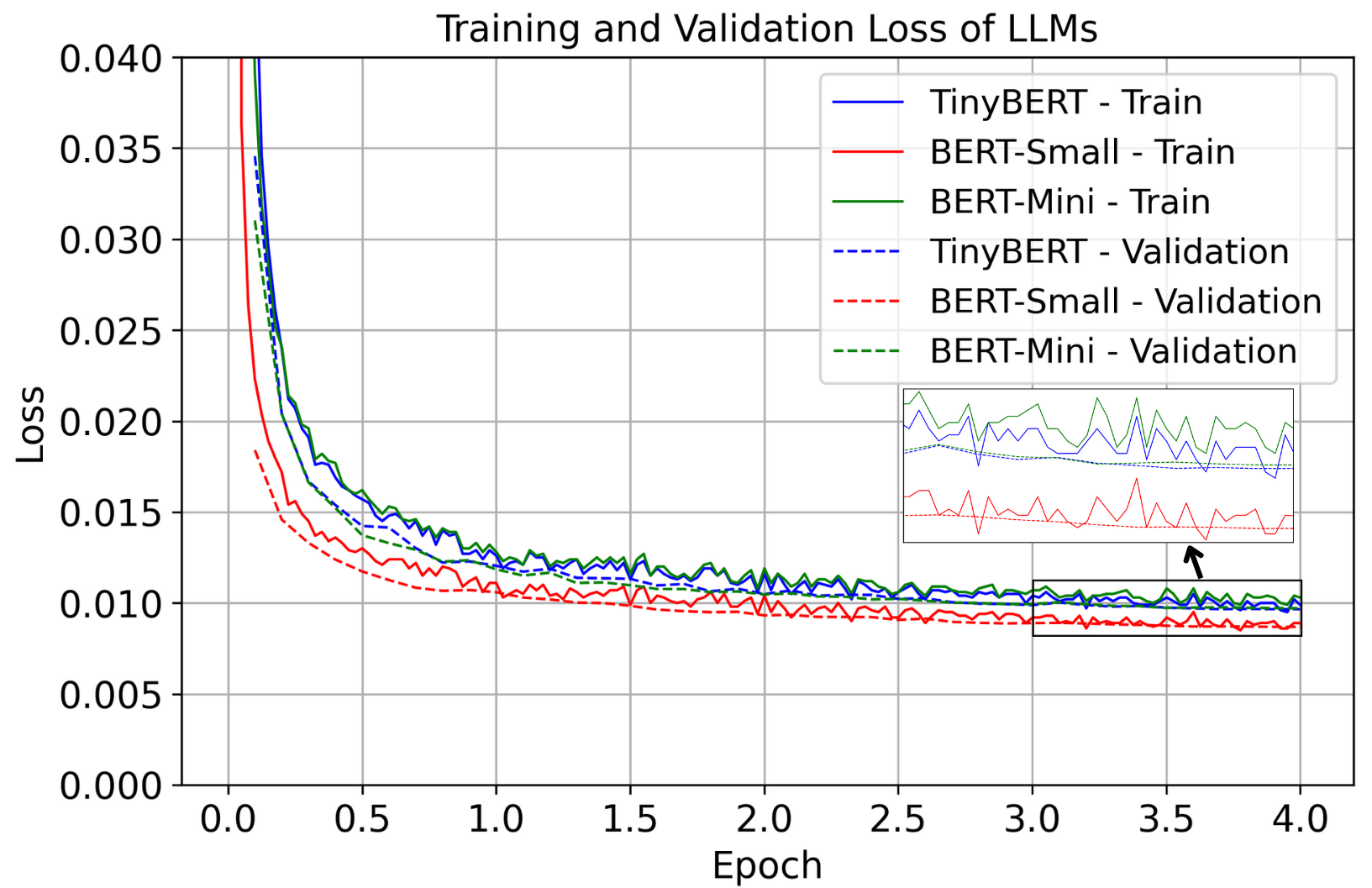} 
    \caption{Training and Validation Loss of LLMs}
    \label{fig:loss}
\end{figure}

\begin{figure*}[t]
    \centering
    \vspace*{0.05in}

    \begin{minipage}{0.32\textwidth}
        \centering
        \includegraphics[width=\linewidth]{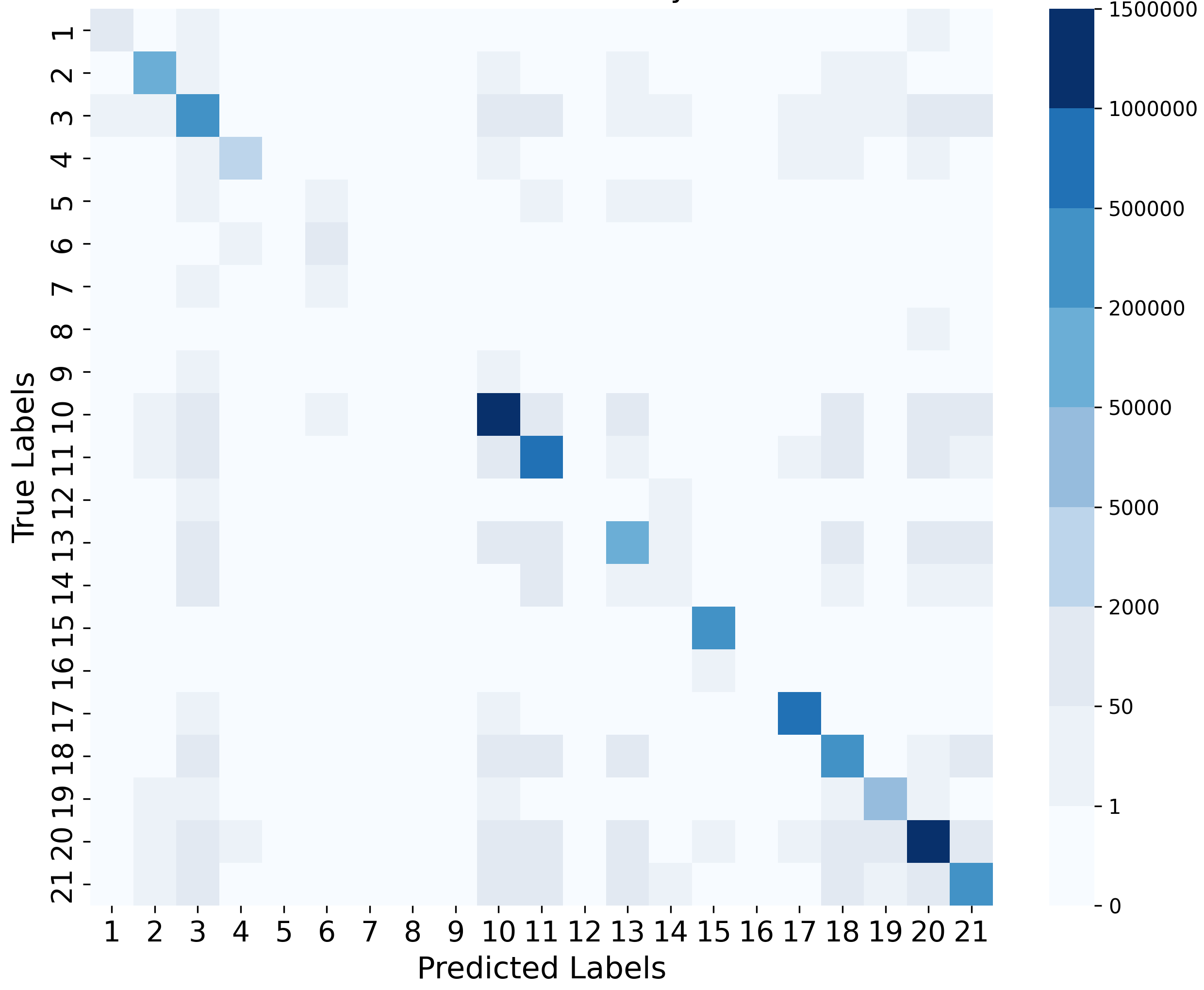}
        \caption{Confusion Matrix of Tiny BERT}
        \label{fig:cmtiny}
    \end{minipage}
    \hfill
    \begin{minipage}{0.32\textwidth}
        \centering
        \includegraphics[width=\linewidth]{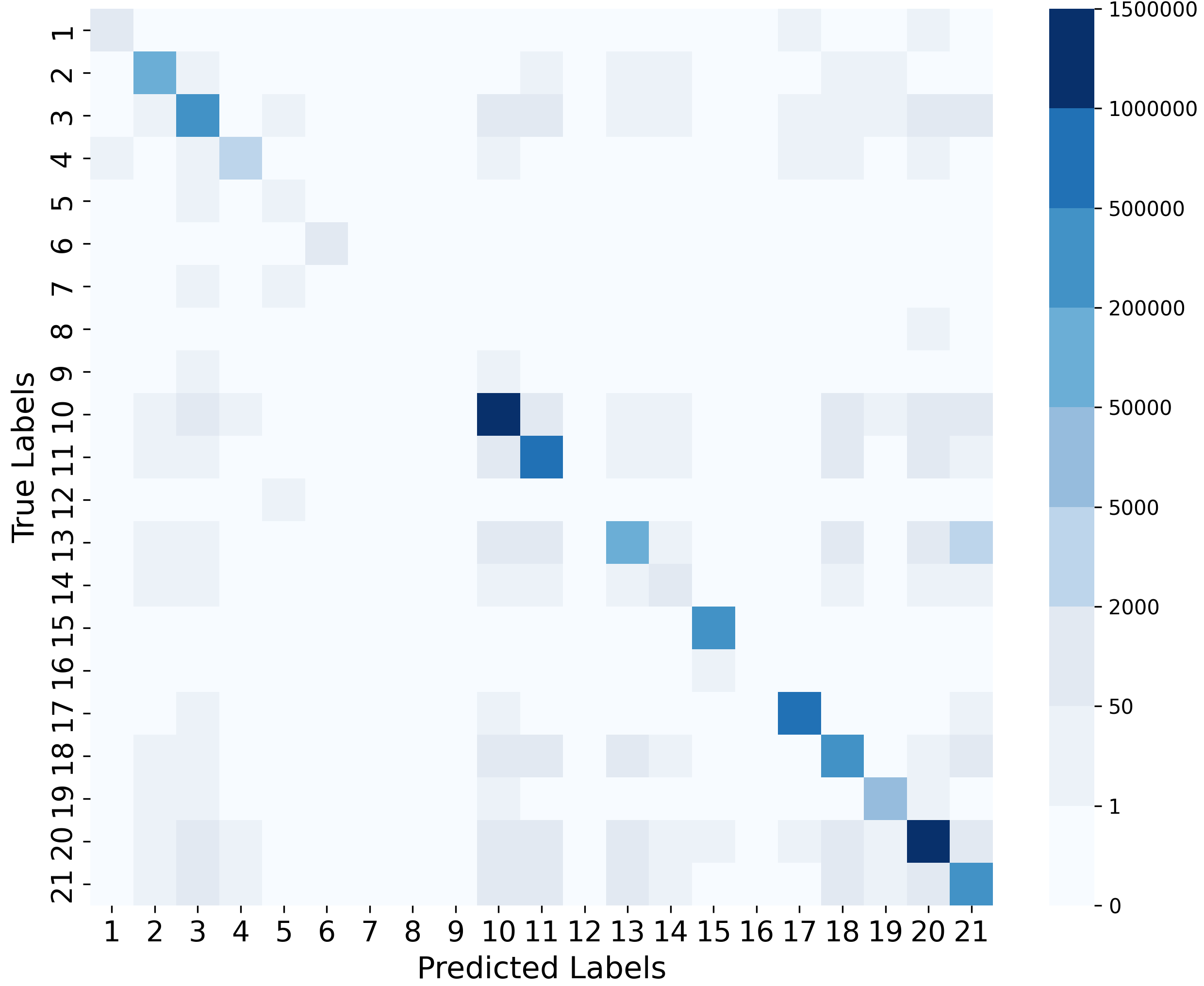}
        \caption{Confusion Matrix of BERT Small}
        \label{fig:cmsmall}
    \end{minipage}
    \hfill
    \begin{minipage}{0.32\textwidth}
        \centering
        \includegraphics[width=\linewidth]{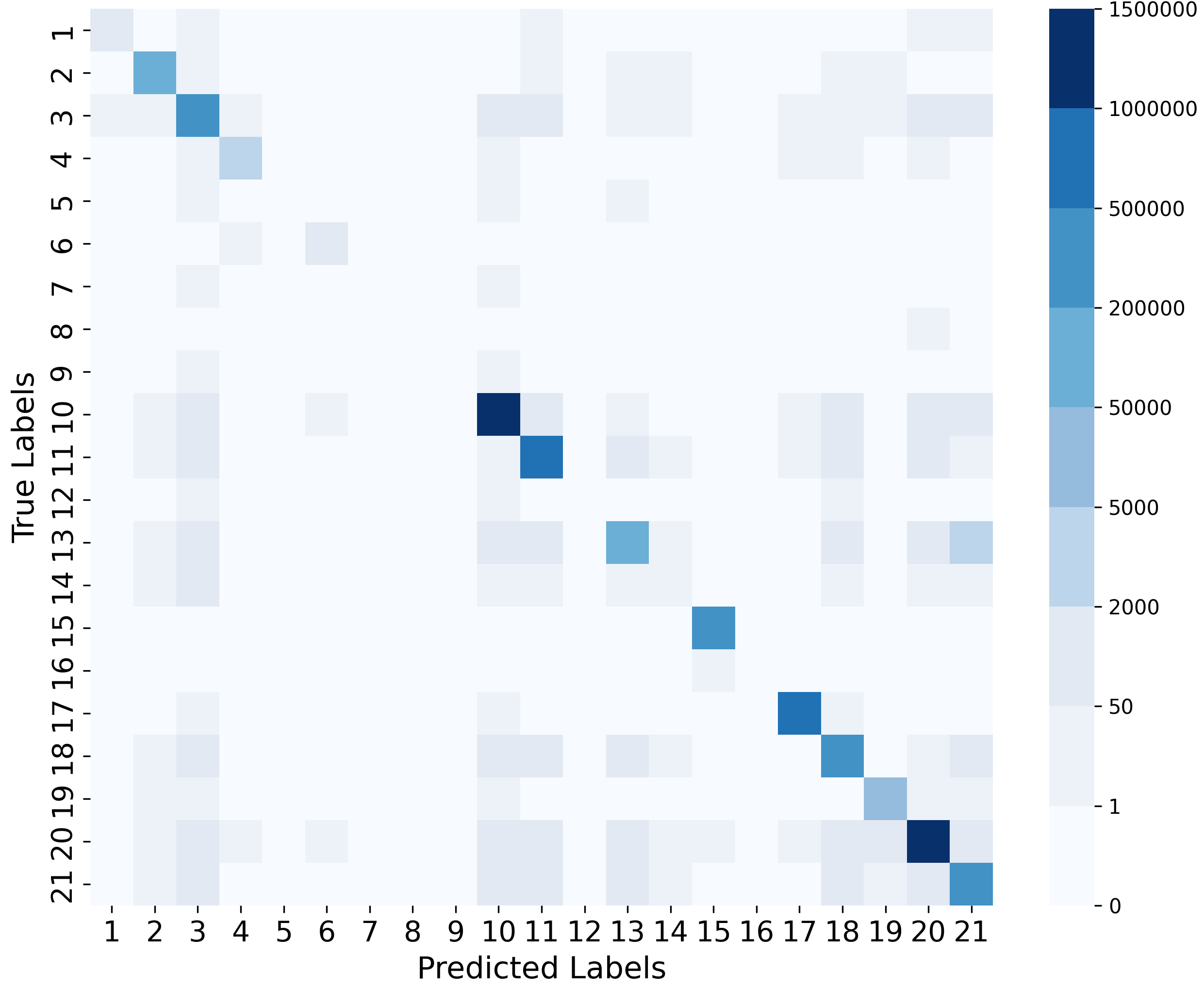}
        \caption{Confusion Matrix of BERT Mini}
        \label{fig:cmmini}
    \end{minipage}
\end{figure*}

\section{Experimental Setup and Results}
\label{sec:Results}

\subsection{Experimental Setup}

A Docker-based environment was developed to emulate the proposed edge--cloud IoT security deployment. Docker containers simulate heterogeneous IoT devices, computationally capable edge nodes, and cloud services while ensuring reproducible and consistent testing. \textit{Edge Layer:} Containers emulate IoT devices, such as smart cameras, industrial sensors, and smart meters, as well as edge nodes, including IoT gateways, edge servers, and local security orchestration nodes. These nodes are used to evaluate edge-level detection and mitigation decision-making. \textit{Cloud Layer:} Dedicated containers host the cloud-based PMM, Global Model Refinement module, and Monitoring Dashboard, supporting prompt and policy updates, system monitoring, and synchronization of logs, action outcomes, and performance indicators. \textit{Dataset and Fine-Tuning:} The IoT-23 and TON\_IoT datasets were merged into a unified corpus and split into 60\% training, 20\% validation, and 20\% testing subsets. The BERT-based detection models were fine-tuned using the HuggingFace \texttt{Trainer} with mixed precision (\texttt{fp16=true}) and the AdamW optimizer. The main hyperparameters were a learning rate of $2\times10^{-5}$, weight decay of $0.01$, $4$ epochs, per-device batch size of $128$, and gradient accumulation of $4$, resulting in an effective batch size of $512$. Cross-entropy loss was used for multi-class classification.

\subsection{Results and Performance Evaluation}

The evaluation focuses on detection performance, inference efficiency, prevention-action accuracy, and resource usage. Fig.~\ref{fig:loss} presents the training and validation loss curves for TinyBERT, BERT-Small, and BERT-Mini over four epochs. The losses decrease progressively across epochs, indicating model convergence, with BERT-Small achieving the lowest training and validation losses. As shown in Table~\ref{tab:results}, all three BERT-based models achieve strong detection performance. BERT Small achieves the best overall results, with 99.75\% test accuracy, F1 Score, precision, and recall, and the lowest training and validation losses of 0.0089 and 0.0086, respectively. Although BERT Mini achieves a slightly higher inference rate of 293.56 requests per second, it has slightly lower accuracy and higher losses than BERT Small. TinyBERT achieves near-equivalent accuracy of 99.73\%, but its inference rate is lower at 232.35 requests per second. These results support selecting BERT Small as the most balanced detection model in terms of predictive performance and inference efficiency. Table~\ref{tab:results_prevention} reports the performance of the Prevention LLM implemented using Mistral-7B-v0.1 and fine-tuned with Low-Rank Adaptation (LoRA) on the custom prevention dataset. The model achieves a training loss of 0.0037, a validation loss of 0.0066, and a test accuracy of 98.3\%, indicating its ability to select suitable mitigation actions across heterogeneous IoT attack scenarios. The fine-tuned Mistral-7B model achieves an average inference throughput of 0.46 requests per second on a single NVIDIA L4 GPU, with an energy consumption of 126.46 J per request. These results indicate that the prevention model is suitable for computationally capable edge infrastructure, such as IoT gateways, edge servers, or local security orchestration nodes. For a model outputting logits \( z_1, z_2, \dots, z_{21} \) for 21 classes, the softmax function converts these logits into probabilities as follows: \( \hat{y}_i = \frac{e^{z_i}}{\sum_{j=1}^{21} e^{z_j}}, \quad \text{for } i = 1, 2, \dots, 21 \), where \( \hat{y}_i \) represents the predicted probability for class \( i \). The cross-entropy loss is defined as \( \mathcal{L}(y, \hat{y}) = -\sum_{c=1}^{21} y_c \log(\hat{y}_c) \), where \( y_c \) is the true label indicator, which equals 1 if class \( c \) is the correct class and 0 otherwise. Micro-F1 aggregates the contributions of all classes by summing the individual true positives, false positives, and false negatives: \( \text{Micro-F1} = \frac{2 \cdot \sum_{i=1}^{21}\text{TP}_i}{2 \cdot \sum_{i=1}^{21}\text{TP}_i + \sum_{i=1}^{21}\text{FP}_i + \sum_{i=1}^{21}\text{FN}_i} \), where \( \text{TP}_i \) denotes the correctly predicted instances of class \( i \), \( \text{FP}_i \) denotes instances incorrectly predicted as class \( i \), and \( \text{FN}_i \) denotes instances of class \( i \) that were not correctly predicted. Figures~\ref{fig:cmtiny}, \ref{fig:cmsmall}, and \ref{fig:cmmini} present the confusion matrices of TinyBERT, BERT Small, and BERT Mini on the test set across 21 classes. BERT Small shows the strongest overall classification behaviour, with a clearer diagonal pattern and fewer off-diagonal confusions compared with the other models. BERT Mini provides competitive performance but shows slightly more misclassifications, while TinyBERT achieves near-equivalent accuracy with lower inference throughput. Across all models, high-frequency classes show stronger classification performance than low-frequency classes, indicating that both model size and label distribution influence the final detection results.

\section{Conclusion}
\label{sec:Conclusion}
The proposed architecture integrates fine-tuned language models for both detection and prevention, leveraging contextual reasoning and adaptability to support anomaly identification and automated mitigation. Trained on IoT-specific datasets, including IoT-23 and TON\_IoT, the framework achieved high detection accuracy across the evaluated lightweight BERT-based models and demonstrated effective prevention performance across diverse attack scenarios. Experimental evaluation in a Docker-based edge–cloud environment confirmed the framework’s ability to detect complex threats while maintaining computational efficiency and scalability. Remaining challenges include optimizing inference latency for resource-constrained devices, enhancing adversarial robustness, and refining deployment for real-time operation. Future work will explore integrating federated and explainable AI techniques to further strengthen scalability, transparency, and decentralized IoT security.

{\footnotesize

\bibliography{bibliography.bib}
\bibliographystyle{IEEEtran}
}
\end{document}